\pdfoutput=1
\documentclass[utf8]{hoarticle}
\setcitestyle{square}
\usepackage{url,hyperref,lineno,microtype,subcaption,pdfpages}
\usepackage[onehalfspacing]{setspace}
\usepackage{amsmath,amssymb,bm}

\def\keyFont{\fontsize{8}{11}\helveticabold }
\def\firstAuthorLast{Villegas and Meloni}
\def\Authors{Pablo Villegas\,$^{1,2,*}$ and Sandro Meloni\,$^{3,1,*}$}
\def\Address{$^{1}$``Enrico Fermi'' Research Center (CREF), Rome, Italy \\
$^{2}$Instituto Carlos I de F\'isica Te\'orica y Computacional, Universidad de Granada, Granada, Spain \\
$^{3}$Institute for Cross-Disciplinary Physics and Complex Systems (IFISC), CSIC-UIB, Palma de Mallorca, Spain}
\def\corrAuthor{pablo.villegas@cref.it, sandro@ifisc.uib-csic.es}
\def\corrEmail{}

\begin{document}
\onecolumn
\firstpage{1}

\title[Higher-order social dilemmas]{Directed percolation in higher-order social dilemmas}

\author[\firstAuthorLast ]{\Authors}
\address{}
\correspondance{}
\extraAuth{}

\maketitle

\begin{abstract}
\section{}
Multiplayer social dilemmas on hypergraphs show a variety of emergent phenomena, which has been attributed to group interactions. Here we test whether pairwise dynamics can reproduce them as well. We derive the mesoscopic Langevin equation of a testbed model and find the operator content of absorbing phase transitions: a directed-percolation line, a tricritical point, and a first-order region whose spinodal is the explosive transition. In our analysis, three-player games only set the values of the Landau coefficients. We construct a strictly pairwise model with short-lived memory, in which every event involves at most two entities, that reproduces these coefficients term by term and has exactly the same mean-field fixed points. Simulations confirm that the two dynamics share mean-field critical and tricritical scaling and the size of the critical mass, and that a purely three-player game on a two-dimensional triangular lattice lies in the directed-percolation class. For these games, the explosive transition and the critical mass therefore do not require group interactions, and the critical behavior is that of the pairwise theory.

\tiny
\keyFont{\section{Keywords:} higher-order interactions, evolutionary game theory, social dilemmas, absorbing phase transitions, directed percolation}
\end{abstract}

\section{Introduction}

How cooperation survives among self-interested agents is a long-standing problem in evolutionary game theory~\citep{Nowak2006,SzaboFath2007,Perc2017}. Its statistical-physics formulation places players on the nodes of a network and lets strategies spread by imitation. Spatial structure then changes the fate of cooperators~\citep{NowakMay1992,SantosPacheco2005,HauertDoebeli2004}. In the last few years, hypergraphs and simplicial complexes have been used to describe interactions among more than two units~\citep{Battiston2020,Battiston2021}. Multiplayer games~\citep{Broom1997,GokhaleTraulsen2010} have accordingly been placed on higher-order structures~\citep{AlvarezRodriguez2021}. In particular, \citet{Civilini2024} showed that, above a critical fraction of three-player games, the Prisoner's Dilemma (PD) undergoes an explosive transition to a bistable state. In that state a cooperative phase coexists with full defection, and a critical mass of initial cooperators is required to reach it. This was interpreted as a novel explanation, rooted in higher-order interactions, for the survival of cooperation~\citep{Civilini2024}. Here we ask a different question: do higher-order games define new collective behavior at large scales?

Universality suggests caution. Very different microscopic rules often lead to the same collective behavior, because coarse-graining erases most of what distinguishes them~\citep{Kadanoff1971,Hinrichsen2000,Odor2004}. Whether group interactions nevertheless change this picture has recently been examined for several canonical models: Ising spins, contagion, voter dynamics, and phase oscillators on hypergraphs~\citep{Robiglio2025,MeloniGabrielliVillegas2026,Llabres2026,SkardalArenas2019}. In each case the group terms modify thresholds, bifurcations, and mechanisms, but not the large-scale behavior, unless they bring in a new symmetry or conservation law~\citep{VillegasMeloni2026}. For instance, simplicial contagion maps onto the Langevin theory of the quadratic contact process, where simplicial terms merely renormalize pairwise coefficients~\citep{MeloniGabrielliVillegas2026}. Here we show that the same holds for evolutionary games. First, the mesoscopic description of higher-order social dilemmas has exactly the operator content of standard absorbing-state field theories. Second, its phase diagram consists of a directed-percolation (DP) line, a tricritical DP (TDP) point, and a first-order region, with the explosive transition as the spinodal of the latter. Third, a strictly pairwise microscopic dynamics with finite-lived memory produces the same field theory. Fourth, simulations of higher-order and pairwise dynamics give the same critical behavior: the pairwise network in mean field, and the ordinary pairwise game in two dimensions.

\section{Higher-order social dilemmas and their field theory}
\label{sec:theory}

\subsection{Model}

\begin{figure}[t!]
\centering
\includegraphics[width=\textwidth]{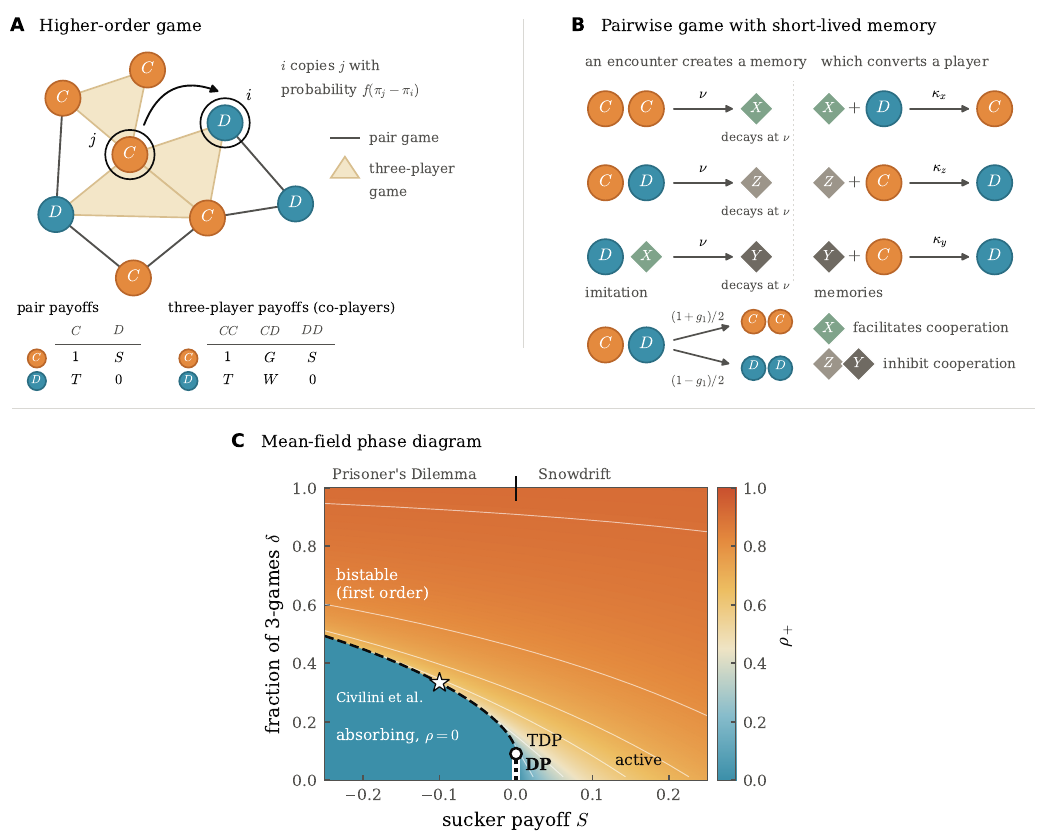}
\caption{Microscopic models and mean-field phase diagram. \textbf{(A)} Higher-order game: players on a hypergraph play pairwise games on 2-hyperedges and three-player games on 3-hyperedges, and a focal player $i$ copies a neighbor $j$ with the Fermi probability $f(\pi_j-\pi_i)$. The tables give the payoffs of a cooperator and of a defector as functions of the strategies of their co-players. \textbf{(B)} Pairwise game with short-lived memory, Eq.~\eqref{eq:network}, in which every event involves at most two entities. Binary encounters create short-lived memories (mediators) that decay at rate $\nu$; the memory $X$ facilitates cooperation, whereas $Z$ and $Y$ inhibit it. Imitation acts at rates $(1\pm g_1)/2$, with a bias $g_1$ toward cooperation. \textbf{(C)} Mean-field phase diagram for $T=1.1$ and $a=1$. The color gives the stable active density $\rho_+$, and thin white lines are its contours. The vertical dashed segment ($S=0$, $\delta<\delta_{\rm tc}$) is the line of continuous DP transitions, and the open circle is the tricritical point (Eq.~\ref{eq:tricritical}). Above it, the absorbing state loses stability at $S=0$ (black tick at the top), while the active branch already exists beyond the spinodal $\delta_{\rm th}(S)$ (black dashed curve), giving a bistable, first-order region. The star marks the explosive transition of~\citep{Civilini2024} ($S=-0.1$, $\delta_{\rm th}=1/3$).}
\label{fig:phase}
\end{figure}

We follow the model of~\citep{Civilini2024} (Figure~\ref{fig:phase}A). Players sit on the nodes of a hypergraph and adopt either cooperation ($C$) or defection ($D$). Each 2-hyperedge hosts a pairwise game with payoffs $1$ (mutual cooperation), $S$ (sucker), $T$ (temptation), and $0$ (mutual defection). Each 3-hyperedge hosts a three-player game with the same $T$ and $S$ and two additional payoffs: a cooperator facing one cooperator and one defector receives $G$, and a defector in the same group receives $W$. The fraction of three-player interactions is $\delta$. At every update a focal player $i$ picks a neighbor $j$ and copies its strategy with the Fermi probability $f(\pi_j-\pi_i)=\{1+\exp[-w(\pi_j-\pi_i)]\}^{-1}$, where $\pi$ denotes payoffs accumulated over all games and $w$ is the selection intensity~\citep{SzaboToke1998,Traulsen2006}. In mean field the dynamics depends only on the product of $w$ and the mean number of games per player, $\langle k\rangle$. Below we denote this product again by $w$, which is set equal to one in~\citep{Civilini2024}. In a well-mixed population with a density $\rho$ of cooperators, the expected payoff differences between a cooperator and a defector in a 2-game and in a 3-game are
\begin{align*}
\Delta\pi_2(\rho)&=\rho(1-T)+(1-\rho)S,\\
\Delta\pi_3(\rho)&=\rho^2(1-T)+2\rho(1-\rho)(G-W)+(1-\rho)^2S,
\end{align*}
and a fraction $\delta$ of 3-games gives $\Delta\pi=(1-\delta)\Delta\pi_2+\delta\,\Delta\pi_3$~\citep{Civilini2024}, that is,
\begin{equation}
\Delta\pi(\rho)=S+(c\delta-b-2S)\rho-c\delta\rho^{2},
\label{eq:dpi}
\end{equation}
where $b=T-S-1$, $a=2(G-W)$, and $c=a+b$. As in~\citep{Civilini2024}, we use $T=1.1$ and $a=1$ unless stated otherwise.

\subsection{Mesoscopic Langevin equation}

Consider coarse-grained cells of $\Omega\gg1$ players. A defector adopts cooperation at rate $W_+=\Omega\rho(1-\rho)f(\Delta\pi)$, and the reverse move occurs at rate $W_-=\Omega\rho(1-\rho)f(-\Delta\pi)$. A Kramers--Moyal expansion of the master equation~\citep{vanKampen2007} gives
\begin{equation}
\partial_t\rho=\rho(1-\rho)\tanh\!\Big[\tfrac{w}{2}\Delta\pi(\rho)\Big]+D\nabla^{2}\rho+\sqrt{\frac{\rho(1-\rho)}{\Omega}}\,\eta(\mathbf{x},t),
\label{eq:langevin}
\end{equation}
where $\eta$ is a Gaussian white noise with $\langle\eta(\mathbf{x},t)\eta(\mathbf{x}',t')\rangle=\delta(\mathbf{x}-\mathbf{x}')\delta(t-t')$ (It\^o convention). The Laplacian stands for the coupling between neighboring cells; on a network it is replaced by $-\sum_jL_{ij}\rho_j$. Equation~\eqref{eq:langevin} is the leading local coarse-grained theory; lattice correlations renormalize its coefficients and generate further terms that are irrelevant at the DP fixed point. Because $f(z)+f(-z)=1$, the noise variance equals $\rho(1-\rho)/\Omega$ for \emph{any} payoff structure. The order of the interaction therefore enters only through the drift.

Expanding Eq.~\eqref{eq:langevin} around the absorbing state $\rho=0$ gives
\begin{equation}
\partial_t\rho=g_1\rho+g_2\rho^{2}-g_3\rho^{3}+D\nabla^{2}\rho+\sigma\sqrt{\rho}\,\eta+\cdots,
\label{eq:landau}
\end{equation}
with $\sigma^2=1/\Omega$ and
\begin{equation}
g_1=\tfrac{w}{2}S,\quad
g_2=\tfrac{w}{2}(c\delta-b-3S),\quad
g_3=\tfrac{w}{2}(2c\delta-b-2S).
\label{eq:coefficients}
\end{equation}
These coefficients follow from the weak-selection drift, $\tanh x\simeq x$. Writing $\Delta\pi=S+u\rho-v\rho^2$ with $u=c\delta-b-2S$ and $v=c\delta$, one has
\begin{equation*}
\tfrac{w}{2}\rho(1-\rho)\Delta\pi=\tfrac{w}{2}\left[S\rho+(u-S)\rho^2-(u+v)\rho^3+v\rho^4\right],
\end{equation*}
and remark (i) below shows that $g_1$, $g_2$, and $g_3$ are exact at $S=0$ for any selection intensity.

Equation~\eqref{eq:landau} is the Reggeon-field-theory representation of Schl\"ogl's models and of the quadratic contact process~\citep{Schlogl1972,GrassbergerTorre1979,Grassberger1982,OhtsukiKeyes1987,Henkel2008}, which is the same equation we obtained for simplicial contagion~\citep{MeloniGabrielliVillegas2026}. It contains a single one-component order-parameter field, a single stable absorbing state, a noise amplitude $\propto\sqrt{\rho}$, and a local polynomial drift. Three-player games add no operator: they only shift $g_2$ and $g_3$. Pairwise games ($\delta=0$) give $g_2<0$, the ordinary contact process. For $c>0$, a sufficient fraction of three-player games drives $g_2$ positive, exactly as facilitation or threshold mechanisms do in pairwise models~\citep{MeloniGabrielliVillegas2026,Bizhani2012,diSanto2018}.

Three remarks sharpen this statement. (i) Since $\tanh x=x+\mathcal O(x^3)$ and $\Delta\pi=\mathcal O(\rho)$ at $S=0$, the exact nonlinear drift reproduces $g_1$, $g_2$, and $g_3$ at $S=0$ for any $w$. Hence the location of the critical line and of the tricritical point does not depend on the selection intensity, consistent with the robustness to $w$ reported in~\citep{Civilini2024}. (ii) The second absorbing state, $\rho=1$, is linearly unstable for $T>1$, since $\partial_t(1-\rho)\simeq\frac{w}{2}(T-1)(1-\rho)$. The payoffs break the $C\leftrightarrow D$ symmetry, so the voter class~\citep{Dornic2001,diSanto2017} is excluded. (iii) An $n$-player game gives $\Delta\pi_n(\rho)=\sum_{k}\binom{n-1}{k}(P^C_k-P^D_k)\rho^k(1-\rho)^{n-1-k}$, where $P^C_k$ and $P^D_k$ are the payoffs of a cooperator and a defector with $k$ cooperating co-players (Supplementary Material, Section~1). Its lowest-order term is $P^C_0-P^D_0$. Unlike simplicial contagion, where $\omega$-simplices enter only at order $\rho^{\omega}$~\citep{MeloniGabrielliVillegas2026}, every multiplayer game contributes to all Landau coefficients. It still generates no new operator, and powers $\rho^{k\geq4}$ are irrelevant by power counting at both the DP and TDP fixed points~\citep{Henkel2008,Lubeck2006}.

\subsection{Phase diagram}

Equations~\eqref{eq:landau} and \eqref{eq:coefficients} organize the whole $(S,\delta)$ plane, shown in Figure~\ref{fig:phase}C. The natural control parameter is the sucker payoff: crossing $S=0$ turns the PD into the Snowdrift game, and $g_1$ changes sign there. For $g_2<0$ the absorbing state loses stability continuously at $S=0$, which is a DP transition. For $g_2>0$ the active branch $\rho_+=[u+(u^{2}+4Sv)^{1/2}]/2v$, is born at finite density, $\rho_+=u/2v$, through a saddle-node bifurcation at the mean-field spinodal
\begin{equation}
\delta_{\rm th}(S)=\frac{b+\sqrt{-4S(b+S)}}{c},\qquad S<0,
\label{eq:spinodal}
\end{equation}
and coexists with $\rho=0$ up to $S=0$, as in Landau--Ginzburg theories of absorbing phase transitions with a positive quadratic term~\citep{Buendia2020SOB}. The two regimes meet at the tricritical point $g_1=g_2=0$,
\begin{equation}
S_{\rm tc}=0,\qquad \delta_{\rm tc}=\frac{T-1}{a+T-1},
\label{eq:tricritical}
\end{equation}
where $\delta_{\rm tc}=1/11$ for the parameters of~\citep{Civilini2024}. The explosive transition reported there is Eq.~\eqref{eq:spinodal} evaluated at $S=-0.1$. In finite dimensions the location of a first-order transition is set by the stalling of fronts between the two phases rather than by the spinodal, but this affects only nonuniversal thresholds. The critical mass of cooperators is the unstable branch $\rho_-$ separating the two basins, exactly as in the quadratic contact process~\citep{OhtsukiKeyes1987,MeloniGabrielliVillegas2026}. Along the line of continuous transitions ($S=0$, $\delta<\delta_{\rm tc}$; vertical dashed segment in Figure~\ref{fig:phase}C) the transition is expected to belong to the DP class, since there is a one-component order parameter with a single stable absorbing state, the interactions are short ranged, and there is no extra symmetry, conservation law, or quenched disorder~\citep{Janssen1981,Grassberger1982,Hinrichsen2000}. At $\delta_{\rm tc}$ it should belong to the TDP class~\citep{OhtsukiKeyes1987,Lubeck2006,Grassberger2006}. 

\section{A pairwise realization with memory}
\label{sec:pairwise}

The analysis above suggests that three-player games enter the collective dynamics only through the values of $g_2$ and $g_3$. If so, any microscopic dynamics that generates the same coefficients should display the same phase diagram, even without group interactions. In a three-player game, the payoff of a player depends simultaneously on two co-players, but a similar dependence can arise from a sequence of pairwise encounters, provided players keep a short-lived memory of them. We make this explicit by constructing a microscopic dynamics in which every event involves at most two entities and that generates the same Landau coefficients as the higher-order game. Besides imitation, $C+D\to C+C$ and $C+D\to D+D$ at rates $(1+g_1)/2$ and $(1-g_1)/2$, where $g_1$ is a bias toward cooperation, the network contains three mediators that decay at rate $\nu$ (Figure~\ref{fig:phase}B):
\begin{align}
C+C&\xrightarrow{\nu}C+C+X, & D+X&\xrightarrow{\kappa_x}C+X,\nonumber\\
C+D&\xrightarrow{\nu}C+D+Z, & C+Z&\xrightarrow{\kappa_z}D+Z,\label{eq:network}\\
D+X&\xrightarrow{\nu}D+X+Y, & C+Y&\xrightarrow{\kappa_y}D+Y.\nonumber
\end{align}
The variable $X$ records encounters between cooperators and recruits defectors. The variables $Z$ and $Y$ record mixed encounters and encounters between a defector and $X$, and both inhibit cooperation. In mean field the mediator densities obey
\begin{equation*}
\dot x=\nu(\rho^2-x),\qquad \dot z=\nu[\rho(1-\rho)-z],\qquad \dot y=\nu[x(1-\rho)-y],
\end{equation*}
so that at stationarity $x=\rho^2$, $z=\rho(1-\rho)$, and $y=\rho^2(1-\rho)$ exactly, for any $\nu$. When the mediators are fast they can be adiabatically eliminated, giving
\begin{equation}
\partial_t\rho=\rho(1-\rho)\left[g_1+(\kappa_x-\kappa_z)\rho-\kappa_y\rho^{2}\right]+\cdots
\end{equation}
with a noise $\sqrt{\rho}$ at leading order. Choosing $g_1=\frac{w}{2}S$, $\kappa_x-\kappa_z=\frac{w}{2}(c\delta-b-2S)$, and $\kappa_y=\frac{w}{2}c\delta$ reproduces Eqs.~\eqref{eq:landau} and \eqref{eq:coefficients} term by term, including the leading noise. All rates are nonnegative for $|S|\leq2/w$ and $c\geq0$, which includes the parameters of~\citep{Civilini2024}. For the parameters used below all eigenvalues of the linearized four-field dynamics at $\rho_+$ have negative real parts; very slow inhibitory memory can instead destabilize $\rho_+$ (Supplementary Material). The quadratic and cubic facilitation that three-player games provide in a single step is here assembled from a sequence of binary encounters. The same structure underlies generalized contagion with memory~\citep{DoddsWatts2004} and the emergence of apparent higher-order effects from pairwise mechanisms in ecology and consensus dynamics~\citep{LettenStouffer2019,Neuhauser2020}.

\section{Numerical tests}
\label{sec:numerics}

We compare both models by simulating the higher-order game and its pairwise counterparts, first in well-mixed populations and on random hypergraphs, where mean-field exponents are exact, and then on a two-dimensional lattice. Details about the numerical simulations are given in the Supplementary Material.

\subsection{Mean field}

\begin{figure}[h!]
\centering
\includegraphics[width=0.72\textwidth]{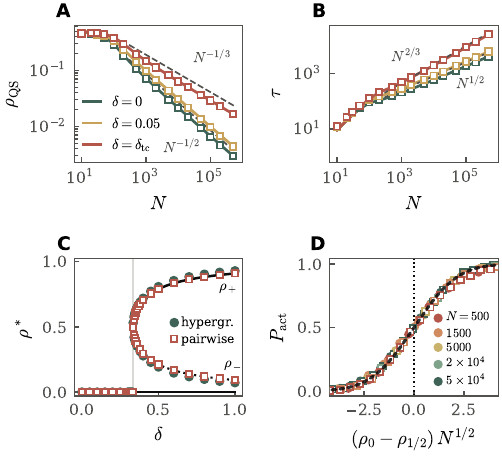}
\caption{Mean field. \textbf{(A)} Quasistationary density and \textbf{(B)} lifetime at $S=0$ versus population size, $10\leq N\leq5\times10^{5}$. Lines: exact QS solution of the well-mixed birth-death process of the game for the pairwise game ($\delta=0$), a higher-order game below the tricritical point ($\delta=0.05$), and the tricritical point ($\delta=\delta_{\rm tc}$). Symbols: Gillespie simulations of the pairwise network of Eq.~\eqref{eq:network} in a well-mixed population, with the same Landau coefficients (error bars smaller than the symbols). Dashed guides: mean-field DP ($\rho\sim N^{-1/2}$, $\tau\sim N^{1/2}$) and TDP ($\rho\sim N^{-1/3}$, $\tau\sim N^{2/3}$). \textbf{(C)} First-order region at $S=-0.1$: stationary densities $\rho^*$ versus $\delta$, with the mean-field branches $\rho_+$ (solid) and $\rho_-$ (dotted). Symbols: agent-based simulations of the higher-order game on random hypergraphs with $\langle k\rangle=20$ (filled circles) and Gillespie simulations of the well-mixed pairwise network (open squares), $N=2\times10^{4}$. Upper symbols: mean of the active QS component. Lower symbols: critical mass $\rho_{1/2}$, the initial density of cooperators from which half of the runs reach the active state. The vertical line marks $\delta_{\rm th}=1/3$. \textbf{(D)} Probability $P_{\rm act}$ of reaching the active state from an initial density $\rho_0$ at $\delta=0.5$, versus $(\rho_0-\rho_{1/2})N^{1/2}$, for the hypergraph game (filled circles) and the pairwise network (open squares), with $N=500$ to $5\times10^{4}$ (brick to dark green). Dotted curve: asymptotic splitting probability $\frac12\{1+\mathrm{erf}[(\rho_0-\rho_{1/2})\sqrt{N\lambda/\rho_-(1-\rho_-)}]\}$, with $\lambda$ the instability rate of $\rho_-$.}
\label{fig:mf}
\end{figure}

At $S=0$ we consider well-mixed populations of $N$ players, $10\leq N\leq5\times10^{5}$. For the game, the Fermi dynamics reduces to a birth-death process for the number of cooperators, and its quasistationary (QS) state~\citep{DickmanVidigal2002,deOliveiraDickman2005} is computed exactly from the principal eigenvector of the generator, without sampling. The pairwise reactions of Eq.~\eqref{eq:network} are simulated in a well-mixed population (equivalently, an annealed homogeneous network) with mass-action rates, using the Gillespie algorithm~\citep{Gillespie1977} and the QS reservoir method~\citep{deOliveiraDickman2005}. Figures~\ref{fig:mf}A,B show that the pairwise game and the higher-order game with $\delta=0.05$ scale as $\rho_{\rm QS}\sim N^{-1/2}$ and $\tau\sim N^{1/2}$, the mean-field DP behavior. At the tricritical point both dynamics cross over to $\rho_{\rm QS}\sim N^{-1/3}$ and $\tau\sim N^{2/3}$, as expected for mean-field TDP~\citep{Lubeck2006}. Both laws follow from balancing a drift $-g\rho^{m}$ with a demographic noise of variance $\rho/N$, which gives $\rho_{\rm QS}\sim N^{-1/m}$ and $\tau\sim N^{(m-1)/m}$, with $m=2$ on the DP line and $m=3$ at the tricritical point~\citep{DickmanVidigal2002,Lubeck2006}.

The pairwise case reproduces not only the exponents but also the amplitudes of the game, with no free parameter, since its rates are fixed term by term by the Landau coefficients of Eq.~\eqref{eq:coefficients}.

Figures~\ref{fig:mf}C,D turn to the first-order region at $S=-0.1$. Here we simulate the higher-order game agent by agent on random hypergraphs built as in~\citep{Civilini2024}, with 2- and 3-hyperedges placed uniformly at random, $\langle k\rangle=20$, and a new hypergraph for each run. The pairwise reactions are simulated, as above, in a well-mixed population of the same size. The active peaks of the QS distributions of both dynamics follow the mean-field branch $\rho_+(\delta)$. The onset of the active state is explosive, with $\rho_+=1/2$ at the spinodal and $\rho_+\simeq0.70$ at $\delta=0.4$. The explosive onset is therefore reproduced without any three-body event.

The same holds for the critical mass. Starting from a random configuration with a fraction $\rho_0$ of cooperators, the probability $P_{\rm act}$ of reaching the active state jumps from $0$ to $1$ at $\rho_{1/2}\simeq\rho_-(\delta)$, the unstable branch of Figure~\ref{fig:mf}C. The width of the jump scales as $N^{-1/2}$, as expected from demographic noise at a saddle point, and the curves of the hypergraph game and of the pairwise network collapse onto one another and onto the parameter-free mean-field prediction (Figure~\ref{fig:mf}D). Unlike the exponents, the critical mass is a nonuniversal quantity, set by the location of the saddle $\rho_-$ and thus by the bare coefficients. Its agreement between the two dynamics is therefore a direct test of the ad-hoc mapping of Eq.~\eqref{eq:coefficients}, rather than a consequence of universality.

\begin{figure}[b!]
\centering
\includegraphics[width=0.85\textwidth]{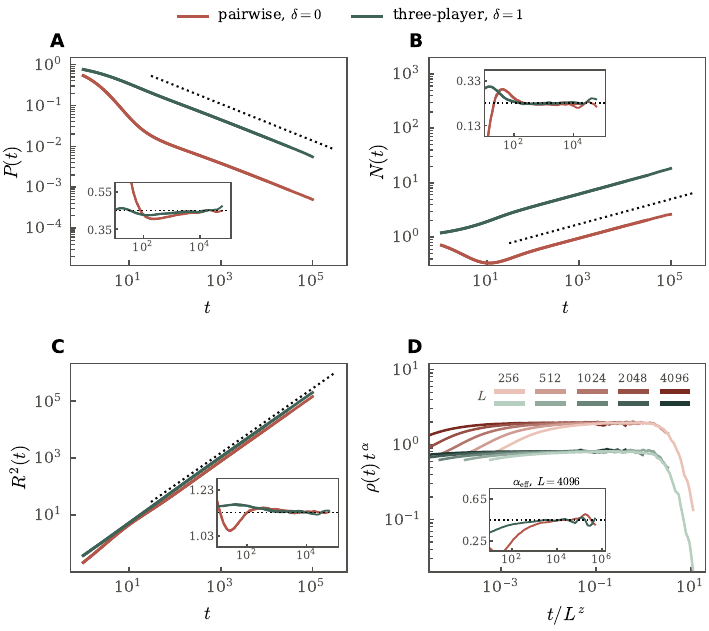}
\caption{Two dimensions. Critical dynamics on triangular lattices for the pairwise game ($\delta=0$) and for a purely three-player game on all elementary triangles ($\delta=1$, $a=-1/2$). \textbf{(A--C)} Spreading from a single cooperator at $S_c^{\rm pw}=0.03754$ and $S_c^{\rm ho}=0.26239$: \textbf{(A)} survival probability $P(t)$, \textbf{(B)} mean number of cooperators $N(t)$, and \textbf{(C)} mean-square radius $R^2(t)$ of surviving clusters. Each curve superimposes three lattice sizes, $L=1024$, $2048$, and $4096$, run up to $t_{\max}=10^4$, $3\times10^4$, and $10^5$, respectively ($4.6\times10^{8}$ pairwise and $4.6\times10^{7}$ three-player runs in total); no cluster reaches the boundary, so the three sizes coincide. Insets: local slopes. \textbf{(D)} Finite-size scaling of the density decay from $\rho_0=1/2$ at $S_c$: $\rho\,t^{\alpha}$ versus $t/L^{z}$ for $L=256$--$4096$ (light to dark), with the 2D DP values of $\alpha$ and $z$. Inset: local slope of $\rho(t)$ for $L=4096$, up to $t=10^{6}$. Dotted lines: 2D DP exponents $\delta_s=\alpha=0.4505$, $\theta=0.2295$, and $2/z=1.1325$~\citep{VoigtZiff1997,GrassbergerZhang1996}.}
\label{fig:lattice}
\end{figure}

\subsection{Two dimensions}

Distinct universality classes can share their mean-field exponents, as DP and conserved DP do. A sharper test therefore requires $d<4$, so we turn to a two-dimensional triangular lattice. We compare a pairwise game played on the edges ($\delta=0$) with a purely higher-order game in which \emph{every} interaction is a three-player game played on the elementary triangles ($\delta=1$). For the latter we take $G=1/4$ and $W=1/2$ ($a=-1/2$), so that $g_2=\frac{w}{2}a<0$ at $S=0$ and the mean-field transition lies on the DP line, away from the tricritical point. For $G=W$ one has $\delta_{\rm tc}=1$, and a purely three-player game would sit exactly at the mean-field tricritical point. 

We measure spreading from a single cooperator in a sea of defectors~\citep{GrassbergerTorre1979,Hinrichsen2000}, which gives $P(t)\sim t^{-\delta_s}$, $N(t)\sim t^{\theta}$, and $R^2(t)\sim t^{2/z}$. For spreading we use lattice sizes, $N=L^2$, with $L=1024$, $2048$, and $4096$, and stop each size before any cluster reaches the boundary, so that the three sizes sample the same infinite-lattice process. This explains the absence of finite-size effects in the spreading curves. We also measure the density decay from a random initial condition, $\rho(t)\sim t^{-\alpha}$, on lattices with $L=256$ to $4096$.

The location of the critical point is nonuniversal and differs widely between the two games: lattice correlations shift it from $S=0$ to $S_c^{\rm pw}=0.03754(1)$ and $S_c^{\rm ho}=0.26239(1)$~\citep{SzaboVukovSzolnoki2005,HauertDoebeli2004}. For both games different sizes overlap, and the local slopes approach the DP values over five decades in time (Figure~\ref{fig:lattice}, insets). Fits over $10^3\leq t\leq10^5$ ($L=4096$) give $\delta_s=0.44(1)$, $\theta=0.22(2)$, and $2/z=1.135(2)$ for the pairwise game, and $\delta_s=0.45(1)$, $\theta=0.23(2)$, and $2/z=1.134(2)$ for the three-player game. The uncertainties reflect the uncertainty in $S_c$. These values agree with 2D DP ($\delta_s=\alpha=0.4505$, $\theta=0.2295$, $2/z=1.1325$)~\citep{VoigtZiff1997,GrassbergerZhang1996}, and they are far from mean-field DP ($\delta_s=\alpha=1$, $\theta=0$, $2/z=1$). The generalized hyperscaling relation $\theta+2\delta_s=d/z$ is thus satisfied for both dynamics (Supplementary Material). The exponents also agree with the known DP behavior of the extinction of cooperators in pairwise lattice games~\citep{SzaboVukovSzolnoki2005}.

The density decay from a random initial condition at $S_c$ gives a consistent picture (Figure~\ref{fig:lattice}D). Fits over $10^{3}\leq t\leq10^{6}$ ($L=4096$) give $\alpha=0.45(3)$ (pairwise) and $0.44(3)$ (three-player). The curves for $L=256$--$4096$ collapse when plotted as $\rho\,t^{\alpha}$ versus $t/L^{z}$ with the DP values of $\alpha$ and $z$, and the smaller lattices leave the critical plateau at $t\simeq L^{z}$ (Figure~\ref{fig:lattice}D); the raw curves are shown in Supplementary Figure~S1. A game with no pairwise interaction at all thus lies in the universality class of the contact process. In two dimensions we have tested only the DP line; the tricritical and first-order regimes are tested in mean field.

\section{Discussion}

Higher-order social dilemmas enrich the set of microscopic mechanisms that produce facilitation. Indeed, their mesoscopic dynamics is the Langevin theory of absorbing phase transitions with one field and one absorbing state. Its phase diagram is the textbook one: a DP line, a TDP point, and a first-order region. Microscopic arity only enters through the bare coefficients of Eq.~\eqref{eq:coefficients}, i.e., through nonuniversal quantities such as thresholds, spinodals, and the size of the critical mass. All of this is consistent with a general principle: arity is not a universality label~\citep{VillegasMeloni2026}. Known exceptions point in the same direction. In the pair-contact process and in parity-conserving branching walks, the order of the reaction matters because it creates additional absorbing states or a conservation law~\citep{Hinrichsen2000,Odor2004}. Likewise, memory with an infinite lifetime, such as permanent immunization, is an additional slow field and changes the class, from DP to dynamic percolation~\citep{Janssen2004,diSanto2017}. The pairwise equivalence above relies on memory that decays at a finite rate.

The same pattern has emerged for other canonical models placed on hypergraphs~\citep{VillegasMeloni2026}. In the Ising model, three-spin terms that reward aligned hyperedges are exact sums of pairwise bonds, and irreducible four-spin terms shift the critical point without changing the mean-field exponents~\citep{Robiglio2025}. In simplicial contagion, 2-simplices add a quadratic term to the Langevin equation of the contact process~\citep{Iacopini2019,MeloniGabrielliVillegas2026}, a term that pairwise facilitation, thresholds, or memory also produce~\citep{DoddsWatts2004,Bizhani2012}. In opinion dynamics, linear group rules reduce to pairwise dynamics on weighted networks~\citep{Neuhauser2020,Llabres2026}, whereas nonlinear rules, such as those of the $q$-voter model, change the collective behavior because they break the conservation of the average opinion~\citep{Neuhauser2020,Castellano2009,Dornic2001,AlHammal2005}. In synchronization, triadic phase interactions among globally coupled oscillators act as a second-harmonic force, and the two-cluster states and explosive transitions they produce also occur in pairwise models~\citep{SkardalArenas2019,Hansel1993,GomezGardenes2011,Pazo2005}. Higher-order social dilemmas are one more instance of this pattern.

Two consequences follow. First, bistability, a finite critical mass, or an explosive onset of cooperation are not signatures of irreducible group interactions. They identify a nonlinear facilitation mechanism~\citep{Buendia2020SOB}, but not its microscopic origin, since a sequence of pairwise encounters with memory produces the same field theory [Eq.~\eqref{eq:network}]~\citep{Rosas2022,Bick2023}. Distinguishing the two requires microscopic information, for instance interventions that separate simultaneous exposure from accumulated memory, or subleading features of the fluctuations. Second, a higher-order game defines genuinely new collective behavior only if its group terms generate a feature that survives coarse graining, such as a new symmetry, a conservation law, an additional slow field, or a long-range coupling. Otherwise, as here, it remains a legitimate and often convenient microscopic choice, and the physics at large scales is fixed by the pairwise field theory.

These results contribute to an ongoing discussion on when higher-order representations are needed~\citep{VillegasMeloni2026,Arenas2026,PerezReche2026,Rommens2026}, and on how the mechanism of a system should be distinguished from its collective behavior. In that language, the present results concern behavior: higher-order and pairwise social dilemmas differ in mechanism, yet share the same large-scale physics.


\section*{Acknowledgments}
S.M. acknowledges the Spanish State Research Agency (MICIU/AEI/10.13039/501100011033) and FEDER (UE) under projects COSASTI (PID2024-157493NB-C22), and the Mar{\'\i}a de Maeztu project CEX2021-001164-M. P.V. acknowledges the Spanish Ministry and Agencia Estatal de Investigaci\'on (AEI), MICIN/AEI/10.13039/501100011033, for financial support, Project PID2023-149174NB-I00 funded also by ERDF/EU.



\bibliographystyle{vancouver}
\bibliography{refs}

\clearpage



\section*{Supplementary material}

# Supplementary Material

## 1 MESOSCOPIC LANGEVIN EQUATION OF HIGHER-ORDER SOCIAL DILEMMAS

### 1.1 Payoffs

In a pairwise game (2-game) the payoff of a player depends on its own strategy and on that of its partner. With the normalization of Civilini et al. [1], mutual cooperation pays $1$, mutual defection pays $0$, a cooperator facing a defector receives $S$, and a defector facing a cooperator receives $T$. In a three-player game (3-game) a cooperator receives $1$, $G$, or $S$ when its two co-players include two, one, or zero cooperators, respectively. A defector receives $T$, $W$, or $0$ in the same three cases. Averaging these payoffs over the binomial composition of the co-players in a well-mixed population gives the payoff differences $\Delta\pi_2$ and $\Delta\pi_3$ quoted in the main text, and hence Eq. (1) of the main text for a fraction $\delta$ of 3-games.

For a general $n$-player game, let $P_k^C$ ($P_k^D$) be the payoff of a cooperator (defector) with $k$ cooperators among its $n-1$ co-players. Then

$$\Delta\pi_n(\rho) = \sum_{k=0}^{n-1} \binom{n-1}{k} (P_k^C - P_k^D)\rho^k(1-\rho)^{n-1-k} \tag{S1}$$

is a polynomial of degree at most $n-1$ whose constant term is $P_0^C - P_0^D$. The payoff of a cooperator among defectors therefore contributes to the linear Landau coefficient for every $n$. This is the main difference with simplicial contagion, where an $\omega$-simplex acts only when $\omega$ of its nodes are active and thus enters at order $\rho^\omega$ [2].

### 1.2 Kramers–Moyal expansion

Consider a mesoscopic cell with $\Omega$ players and $n$ cooperators, $\rho = n/\Omega$. In the pairwise-comparison (Fermi) process a focal player copies a randomly chosen partner with probability $f(\pi_{\text{partner}} - \pi_{\text{focal}})$, with $f(z) = [1 + e^{-wz}]^{-1}$. Measuring time in Monte Carlo steps ($\Omega$ updates), the rates at which $n$ increases or decreases by one are

$$W_\pm(n) = \Omega\,\rho(1-\rho)\,f(\pm\Delta\pi). \tag{S2}$$

The first two jump moments are

$$A(\rho) = \frac{W_+ - W_-}{\Omega} = \rho(1-\rho)\,[f(\Delta\pi) - f(-\Delta\pi)] = \rho(1-\rho)\tanh\!\left(\frac{w}{2}\Delta\pi\right), \tag{S3}$$

$$B(\rho) = \frac{W_+ + W_-}{\Omega^2} = \frac{\rho(1-\rho)}{\Omega}\,[f(\Delta\pi) + f(-\Delta\pi)] = \frac{\rho(1-\rho)}{\Omega}. \tag{S4}$$

The second identity uses $f(z) + f(-z) = 1$ and is exact. Truncating the expansion at second order and coupling neighboring cells through imitation gives Eq. (2) of the main text, with Itô multiplicative noise. Here $w$ denotes the selection intensity per game. When payoffs are accumulated over $\langle k\rangle$ games and the Fermi rule acts on total payoffs with intensity $w_{\text{tot}}$, one has $w = w_{\text{tot}}\langle k\rangle$. The original model [1] sets $w_{\text{tot}} = 1/\langle k\rangle$, i.e., $w = 1$.

### 1.3 Landau coefficients

The weak-selection expansion of the drift, $A(\rho) \simeq \frac{w}{2}\rho(1-\rho)\Delta\pi$, is given in the main text and yields Eqs. (3) and (4); we checked it symbolically. Beyond weak selection, $\tanh x = x - x^3/3 + \ldots$ generates corrections of order $\Delta\pi^3$. At $S = 0$, $\Delta\pi = \mathcal{O}(\rho)$, so these corrections are $\mathcal{O}(\rho^4)$ and leave the Landau coefficients $g_1$, $g_2$, and $g_3$ unchanged (the leading correction is $-w^3u^3\rho^4/24$). When, in addition, $u = 0$, they are $\mathcal{O}(\rho^7)$. The critical line $S = 0$ and the tricritical point $\delta_{\rm tc}$ are therefore exact for any selection intensity. Near $\rho = 1$, setting $\psi = 1 - \rho$ gives $\partial_t\psi \simeq -\frac{w}{2}\Delta\pi(1)\psi = \frac{w}{2}(T-1)\psi$. Full cooperation is also absorbing under pure imitation, but it is linearly unstable whenever $T > 1$, so the transition at $\rho = 0$ involves a single stable absorbing state.

## 2 MEAN-FIELD PHASE DIAGRAM

With $\Delta\pi = S + u\rho - v\rho^2$, $u = c\delta - b - 2S$, and $v = c\delta$, the stationary states of the weak-selection drift are $\rho = 0$, $\rho = 1$, and the roots of $S + u\rho - v\rho^2 = 0$,

$$\rho_\pm = \frac{u \pm \sqrt{u^2 + 4Sv}}{2v}. \tag{S5}$$

(i) For $S > 0$ (and $T > 1$, so that $\Delta\pi(1) = 1 - T < 0$) there is a single interior root $\rho_+$, which is stable, while $\rho = 0$ is unstable. (ii) For $S < 0$ the absorbing state is stable. Interior roots exist when $u > 0$ and $u^2 + 4Sv \geq 0$, which gives the spinodal of Eq. (5) of the main text (Eq. (4) of [1]). There $\rho_+$ is stable and $\rho_-$ is unstable. (iii) As $S \to 0^+$, $\rho_+ \to u/v$ if $u > 0$ and $\rho_+ \to 0$ if $u < 0$. The transition at $S = 0$ is thus continuous for $u < 0$ and discontinuous for $u > 0$. At $S = 0$ one has $u = c\delta - (T-1)$, and the change of regime occurs at the tricritical point $\delta_{\rm tc} = (T-1)/(a+T-1)$, which coincides with $\lim_{S\to 0^-}\delta_{\rm th}(S)$. For the parameters used in the main text ($T = 1.1$, $a = 1$), $\delta_{\rm tc} = 1/11 \simeq 0.091$.

## 3 PAIRWISE REALIZATION WITH MEMORY

### 3.1 Reactions

The pairwise reactions of Eq. (7) in the main text contains, besides $C$ and $D$ (with $N_C + N_D = N$ conserved), three mediator species $X$, $Y$, and $Z$ whose numbers are not conserved. With mass-action propensities (densities $x = N_X/N$, etc.), the reactions are:

| reaction | propensity | effect |
|---|---|---|
| $C + D \to D + D$ | $\frac{N}{2}(1-g_1)\rho(1-\rho)$ | $\rho \to \rho - 1/N$ |
| $C + D \to C + C$ | $\frac{N}{2}(1+g_1)\rho(1-\rho)$ | $\rho \to \rho + 1/N$ |
| $C + C \to C + C + X$ | $N\nu\rho^2$ | $x \to x + 1/N$ |
| $X \to \varnothing$ | $N\nu x$ | $x \to x - 1/N$ |
| $D + X \to C + X$ | $N\kappa_x x(1-\rho)$ | $\rho \to \rho + 1/N$ |
| $C + D \to C + D + Z$ | $N\nu\rho(1-\rho)$ | $z \to z + 1/N$ |
| $Z \to \varnothing$ | $N\nu z$ | $z \to z - 1/N$ |
| $C + Z \to D + Z$ | $N\kappa_z z\rho$ | $\rho \to \rho - 1/N$ |
| $D + X \to D + X + Y$ | $N\nu x(1-\rho)$ | $y \to y + 1/N$ |
| $Y \to \varnothing$ | $N\nu y$ | $y \to y - 1/N$ |
| $C + Y \to D + Y$ | $N\kappa_y y\rho$ | $\rho \to \rho - 1/N$ |

Every event involves at most two entities, and no event involves three players simultaneously. The mean-field equations are

$$\dot{\rho} = g_1\rho(1-\rho) + \kappa_x x(1-\rho) - \kappa_z z\rho - \kappa_y y\rho, \tag{S6}$$

$$\dot{x} = \nu(\rho^2 - x), \qquad \dot{z} = \nu[\rho(1-\rho) - z], \qquad \dot{y} = \nu[x(1-\rho) - y]. \tag{S7}$$

The stationary mediator densities are $x^* = \rho^2$, $z^* = \rho(1-\rho)$, and $y^* = \rho^2(1-\rho)$ exactly. The fixed points in $\rho$ therefore coincide with those of

$$\dot{\rho} = \rho(1-\rho)\left[g_1 + (\kappa_x - \kappa_z)\rho - \kappa_y\rho^2\right] \tag{S8}$$

for any $\nu$. The same equation describes the transient dynamics in the fast-mediator limit $\nu \to \infty$ at fixed $\kappa$'s. The identification with the game is $g_1 = \frac{w}{2}S$, $\kappa_x - \kappa_z = \frac{w}{2}u$ (we use $\kappa_x = \frac{w}{2}\max(u,0)$ and $\kappa_z = \frac{w}{2}\max(-u,0)$), and $\kappa_y = \frac{w}{2}v$.

# 4 NUMERICAL METHODS

## 4.1 Well-mixed populations and hypergraphs

For a population of $N$ players the Fermi process is a birth-death chain with rates $W_\pm(n) = N\frac{n(N-n)}{N(N-1)}f[\pm\Delta\pi(n/N)]$. Its QS distribution and lifetime follow from the principal eigenvector and eigenvalue of the generator restricted to $n = 1,\ldots,N-1$ [3]. Near the saddle $\rho_-$ the drift is $A(\rho) \simeq \lambda(\rho-\rho_-)$, with $\lambda = A'(\rho_-) > 0$, and the noise variance is $\rho_-(1-\rho_-)/N$. The backward equation then gives the splitting probability $P_{\text{act}} = \frac{1}{2}[1 + \text{erf}((\rho_0 - \rho_-)\sqrt{N\lambda/\rho_-(1-\rho_-)})]$, the curve of Figure 2D of the main text; at $\delta = 0.5$ it agrees with the exact splitting probability of the chain for $N = 2\times10^4$. The pairwise network is simulated with the Gillespie algorithm [4] ($\nu = 10$) and the QS reservoir method [5], with a reservoir memory much longer than $\tau$; faster mediators ($\nu = 30, 100$) give the same results. Random hypergraphs are built as in [1], with $\langle k\rangle = 20$ and $w_{\text{tot}} = 1/\langle k\rangle$, and each run involves independent realizations. For the critical mass, runs start from a random configuration with density $\rho_0$ and stop at extinction or at $\rho_s = (\rho_- + 2\rho_+)/3$; $P_{\text{act}}$ is the fraction of runs that reach $\rho_s$, estimated from $500$–$2000$ runs per point.

## 4.2 Triangular lattice

Players occupy an $L\times L$ triangular lattice with periodic boundaries. In the pairwise game every edge hosts a 2-game; in the higher-order game every elementary triangle hosts a 3-game ($G = 1/4$, $W = 1/2$). Players imitate nearest neighbors with the Fermi rule applied to total payoffs ($T = 1.1$, $w_{\text{tot}} = 1$), with continuous-time random-sequential updates. Spreading runs start from a single cooperator [6] on lattices with $L = 1024$, $2048$, and $4096$, up to $t = 10^4$, $3\times10^4$, and $10^5$, respectively, so that no cluster reaches the boundary ($\sim 10^7$ runs per value of $S$). The critical point is the value of $S$ at which the late-time curvature of $\ln P$ and $\ln N$ versus $\ln t$ vanishes, with $\ln P$ and $\ln N$ interpolated in $S$ between five simulated values; it is the same for the three sizes. Exponents are fitted over the last two decades at $S_c$, and their uncertainties are the changes for $S_c \pm 10^{-5}$. At $L = 4096$ the hyperscaling combination $\theta + 2\delta_s$ holds for both cases.

## 4.3 Density decay

Starting from a random configuration with $\rho_0 = 1/2$, we follow $\rho(t)$ for $L = 256$–$2048$ up to $t = 2\times10^5$ (1000 to 100 realizations) and for $L = 4096$ up to $t = 10^6$ (40 realizations). At a DP critical point

$\rho(t) \sim t^{-\alpha}$ with $\alpha = \delta_s$ by rapidity reversal, and $\rho(t, L) = t^{-\alpha} F(t/L^z)$ on a finite lattice. Each size is run at three values of $S$ spaced by $3 \times 10^{-5}$ around $S_c$, and $\ln \rho$ is interpolated in $S$ as for spreading.

The density decay reaches later times than spreading and therefore provides an independent test of the critical point. Away from $S_c$ the effective exponent bends toward extinction or toward the active phase, whereas at $S_c$ it settles on the DP value (Supplementary Figure S1C,D). For all sizes the raw curves follow the same power law until $t \simeq L^z$ (Supplementary Figure S1A,B), which gives the collapse of Figure 3D of the main text.

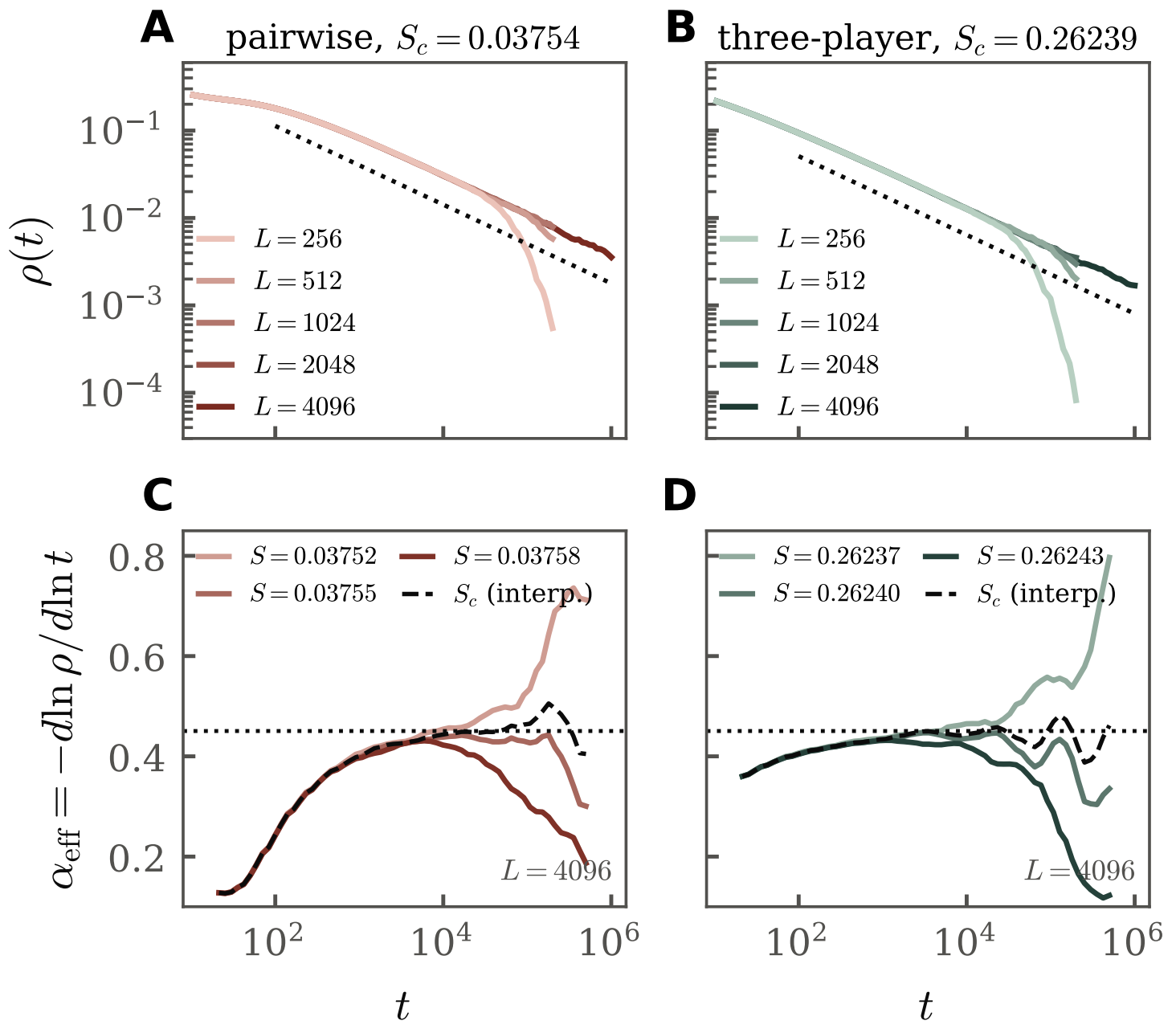


**Figure S1.** Density decay from $\rho_0 = 1/2$ on the triangular lattice. **(A,B)** $\rho(t)$ at $S_c$ for $L = 256$–$4096$ (light to dark); dotted lines: $t^{-0.4505}$. **(C,D)** Effective exponent $\alpha_{\text{eff}} = -d\ln\rho/d\ln t$ for $L = 4096$ at the three simulated values of $S$ (solid, light to dark with increasing $S$) and at the interpolated $S_c$ (dashed); dotted lines: $\alpha = 0.4505$.



\end{document}